\documentclass[aps,prb,twocolumn,superscriptaddress,amsmath]{revtex4-1}

\usepackage{graphicx}
\usepackage{color}

\begin{document}

%\preprint{}

\title{Spin-polarized quasiparticle transport in exchange-split superconducting aluminum on europium sulfide}


\author{M. J. Wolf}
\affiliation{Institut f\"ur Nanotechnologie, Karlsruher Institut f\"ur Technologie (KIT), Karlsruhe, Germany}
\author{C. S\"urgers}
\author{G. Fischer}
\affiliation{Physikalisches Institut, Karlsruher Institut f\"ur Technologie (KIT), Karlsruhe, Germany}
\author{D. Beckmann}
\email[e-mail address: ]{detlef.beckmann@kit.edu}
\affiliation{Institut f\"ur Nanotechnologie, Karlsruher Institut f\"ur Technologie (KIT), Karlsruhe, Germany}

\date{\today}

\begin{abstract}
We report on nonlocal spin transport in mesoscopic superconducting aluminum wires in contact with the ferromagnetic insulator europium sulfide. We find spin injection and long-range spin transport in the regime of the exchange splitting induced by europium sulfide. Our results demonstrate that spin transport in superconductors can be manipulated by ferromagnetic insulators, and opens a new path to control spin currents in superconductors.
\end{abstract}

\pacs{72.25.-b,74.25.F-,74.45.+c,74.78.Na,85.75.-d}
% 72.25.-b 	Spin polarized transport 
% 73.23.-b 	Electronic transport in mesoscopic systems
% 73.40.Gk = Tunneling
% 74.25.F- = Superconductors / Transport properties
% 74.40.Gh = Nonequilibrium superconductivity
% 74.45.+c 	Proximity effects; Andreev reflection; SN and SNS junctions
% 74.55.+v 	Tunneling phenomena: single particle tunneling and STM
% 74.78.Na 	Mesoscopic and nanoscale systems 
% 85.75.-d 	Magnetoelectronics; spintronics: devices exploiting spin polarized transport or integrated magnetic fields
%\keywords{}

\maketitle


\section{Introduction}

In conventional superconductors, electrons are bound in singlet Cooper pairs with zero spin. In hybrid structures with magnetic elements, triplet Cooper pairs and spin-polarized supercurrents can be created.\cite{bergeret2001,bergeret2005,keizer2006,khaire2010,robinson2010,eschrig2011} In addition, spin-polarized quasiparticles can be injected into superconductors,\cite{johnson1994} with very long spin relaxation times.\cite{yang2010,huebler2012b,quay2013} Both effects open the possibility to achieve spintronics functionality with superconductors. Recently, we have observed long-range spin-polarized quasiparticle transport in superconducting aluminum in the presence of a large Zeeman splitting of the density of states,\cite{huebler2012b} and demonstrated that the superconductor acts as a spin filter for quasiparticles injected from a paramagnetic metal.\cite{wolf2013} These observations open the possibility to use superconducting aluminum as an active element for spintronics. The design of complex structures with switchable elements requires, however, a local control of the spin splitting, which cannot be achieved by a homogeneous applied magnetic field. In hybrid structures of ferromagnetic insulators and superconductors, an exchange splitting of the density of states of the superconductor can be induced due to spin-active scattering at the interface between the materials.\cite{tokuyasu1988,millis1988} Such structures are candidates for new spintronics functionality in superconductors,\cite{huertas-hernando2002,giazotto2008,li2013,pal2014} and in particular to control and manipulate the spin splitting.\cite{li2013b} A very promising material for this purpose is europium sulfide (EuS), which has been used in the past in superconducting hybrid structures as a means to induce an exchange splitting,\cite{hao1991,xiong2011} as a spin-filter material,\cite{moodera1988,hao1990} and for superconducting spin valves\cite{li2013} and spin switches.\cite{li2013b} EuS is a II-VI semiconductor with a direct band gap of $E_g = 1.7~\mathrm{eV}$ at room temperature and exhibits isotropic Heisenberg ferromagnetism with a Curie temperature $T_\mathrm{C} = 16.5~\mathrm{K}$.\cite{zinn1976} At low temperatures (few K) it can be considered as an insulator. In this paper, we extend our previous experiments\cite{huebler2012b,wolf2013} to spin transport in mesoscopic superconducting aluminum wires with an exchange splitting induced by the ferromagnetic insulator EuS. 

\section{Samples and Experiment}

\begin{figure}[bt]
\includegraphics[width=\columnwidth]{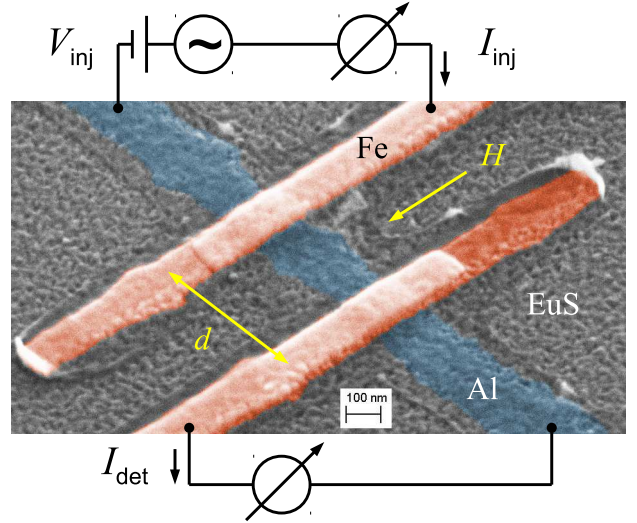}
\caption{\label{fig_sample}
(color online) False-color scanning electron microscopy image of a section of one of our samples together with the measurement scheme with the injection (inj) and detection (det) circuits.}
\end{figure}

\begin{table*}[bt] %add [H] placement to break table across pages
\caption{\label{tab_sample_properties}Overview of sample properties. Range of normal-state tunnel conductances $G$, EuS film thickness $t_\mathrm{EuS}$. Aluminum film properties: film thickness $t_\textrm{Al}$, critical temperature $T_c$, critical magnetic field $\mu_0 H_\mathrm{c}$, diffusion constant $D$, maximum exchange field  $B^*_\mathrm{sat}$.}
\begin{ruledtabular}
%\begin{tabular}{l p{1.5cm} *{2}{p{1.0cm}} *{2}{p{1.75cm}}*{4}{p{1.0cm}}p{1.5cm}}
\begin{tabular}{llccccccccc}
        & $G$          & $t_\mathrm{EuS}$ & $t_\mathrm{Al}$ & $T_c$  & $\mu_0 H_\mathrm{c}$ & $D$   & $B^*_\mathrm{sat}$  \\
Sample  & ($\mu$S)     & (nm)             & (nm)            & (K)    & (T)    & ($\mathrm{cm^2/s}$) & (T)  \\ \hline
A       &  $690-750$   & 22               & 10.5            & 1.55   & 0.95                 & 20.4  & 1.75 \\
B       &  $380-450$   & 10               & 9.5             & 1.6    & 1.45                 & 13.6  & 1.2 
\end{tabular}
\end{ruledtabular}
\end{table*}

Our samples were fabricated in a two-step process. First, EuS films of thickness $t_\mathrm{EuS}=10-25~\mathrm{nm}$ were evaporated onto a Si (111) substrate heated to $T_\mathrm{S}\approx800~^\circ \mathrm{C}$. The films have a strong $\langle 111\rangle$ texture with a small fraction of $\langle 100\rangle$-oriented grains. The Curie temperature $T_\mathrm{C}\approx 16.4~\mathrm{K}$ and saturation magnetization $M_\mathrm{s}\approx 7~\mu_\mathrm{B}$ per formula unit agree well with bulk properties. The coercive field is typically a few mT, consistent with negligible magnetocrystalline anisotropy as expected for a Heisenberg ferromagnet. Details of the film preparation and characterization can be found in Ref.~\onlinecite{wolf2014b}. In the second step, aluminum/iron structures were fabricated by e-beam lithography and shadow evaporation techniques on top of the EuS films. The EuS films were coated with PMMA resist, and after exposure and development they were mounted in the evaporation chamber. First, a short Ar ion etching step was used to clean the exposed surface of the EuS film to ensure good contact with the metal films. Next, a thin superconducting aluminum strip of thickness $t_\mathrm{Al}\approx 10~\mathrm{nm}$ was evaporated and then oxidized \textit{in situ} to form an insulating tunnel barrier. Then ferromagnetic iron ($t_\mathrm{Fe}\approx15-25~\mathrm{nm}$) was evaporated under a different angle to form six tunnel junctions to the aluminum, with contact separations $d$ spanning $0.5$ to $5~\mathrm{\mu m}$. Additional copper layers were evaporated under different angles to reduce the resistance of the iron leads.  Figure~\ref{fig_sample} shows a scanning electron microscopy image of a section of one of our samples, together with the experimental scheme.

The samples were mounted into a shielded box thermally anchored to the mixing chamber of a dilution refrigerator. A magnetic field was applied in the plane of the substrate, along the direction of the iron wires. We will refer to the applied field by $H$ throughout this paper, whereas $B$ will be used to describe the effective spin splitting of the density of states (both field-induced and exchange-induced). Using a combination of dc bias and low-frequency ac excitation, we measured both the local differential conductance $g_\mathrm{loc}=dI_\mathrm{inj}/dV_\mathrm{inj}$ of individual junctions as well as the nonlocal differential conductance $g_\mathrm{nl}=dI_\mathrm{det}/dV_\mathrm{inj}$ for different injector/detector pairs in the superconducting state. For the nonlocal conductance, we plot the normalized signal $\hat{g}_\mathrm{nl}=g_\mathrm{nl}/\left(G_\mathrm{inj}G_\mathrm{det}\right)$ throughout the paper, where $G_\mathrm{inj}$ and $G_\mathrm{det}$ are the normal-state conductances of the injector and detector junction, respectively. Also, we mainly focus on data at $T=50~\mathrm{mK}$. In addition to the conductances in the superconducting state, we measured the nonlocal linear resistance $R_\mathrm{nl}=dV_\mathrm{det}/dI_\mathrm{inj}$ in the normal state at $T=4.2~\mathrm{K}$.

Similar results were obtained on two samples (A and B). An overview of sample parameters can be found in Table~\ref{tab_sample_properties}. We will focus here mostly on results from sample A.

\section{Results}

\begin{figure}[bt]
\includegraphics[width=\columnwidth]{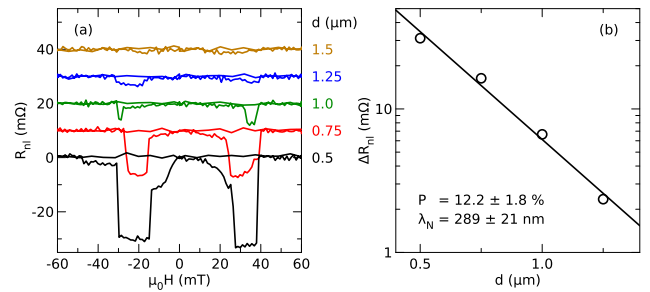}
\caption{\label{fig_spinvalve}
(color online) (a) Nonlocal resistance $R_\mathrm{nl}$ as a function of applied magnetic field $H$ for different contact separation $d$ in the normal state at $T=4.2~\mathrm{K}$. The data are shifted vertically for clarity. (b) Spin-valve signal $\Delta R_\mathrm{nl}$ as a function of contact distance $d$ (symbols), and fit to an exponential decay (line).}
\end{figure}

Before we describe the results in the superconducting state, we first characterize spin transport in the normal state. Figure \ref{fig_spinvalve}(a) shows the nonlocal resistance $R_\mathrm{nl}$ as a function of applied magnetic field $H$ for different pairs of contacts at $T=4.2~\mathrm{K}$. Data are shown for major hysteresis loops in both sweep directions, and offset for clarity. In both sweep directions, two switches can be observed between parallel (P) and antiparallel (AP) configuration. From these, we extract the spin-valve signal $\Delta R_\mathrm{nl}=R_\mathrm{nl}^\mathrm{(P)}-R_\mathrm{nl}^\mathrm{(AP)}$. The spin-valve signal is plotted as a function of contact distance $d$ in Fig.~\ref{fig_spinvalve}(b), together with a fit to the standard expression\cite{valet1993,jedema2002}
\begin{equation}
\Delta R_\mathrm{nl}=P^2\frac{\rho\lambda_\mathrm{N}}{\mathcal{A}}\exp(-d/\lambda_\mathrm{N}),
\end{equation}
where $P$ is the spin polarization of the tunnel conductance, $\rho$ is the normal-state resistivity of the aluminum, $\mathcal{A}$ is the cross-section area of the aluminum, and $\lambda_\mathrm{N}$ is the normal-state spin-diffusion length. From the fit, we obtain $P=12.2\pm1.8\%$ and $\lambda_\mathrm{N}=289\pm21~\mathrm{nm}$. With the electron diffusion coefficient $D$ determined from the normal-state resistivity we obtain the spin relaxation time $\tau_\mathrm{sf}=\lambda^2_\mathrm{N}/D=41~\mathrm{ps}$.

\begin{figure}[bt]
\includegraphics[width=\columnwidth]{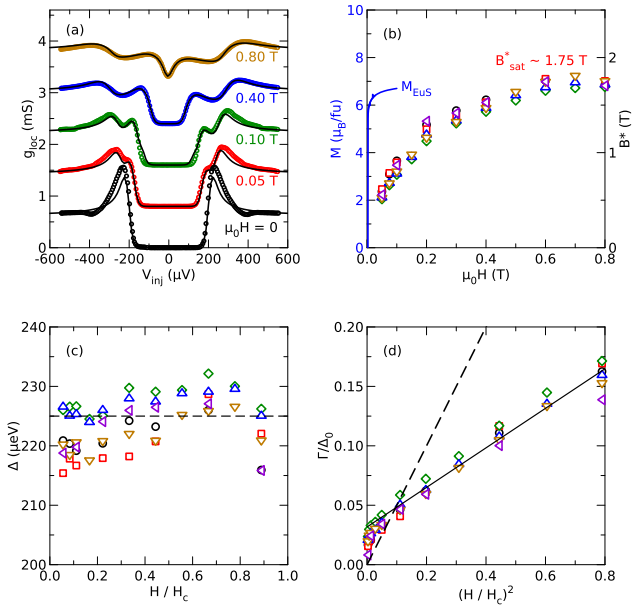}
\caption{\label{fig_local} 
(color online) (a) Local differential conductance $g_\mathrm{loc}=dI_\mathrm{inj}/dV_\mathrm{inj}$ of one junction as a function of injector bias $V_\mathrm{inj}$ for different applied magnetic fields $H$ (symbols) and fits (lines). 
(b) Magnetization of the EuS film (solid line, left ordinate) and induced exchange field $B^*$ for different contacts (symbols, right ordinate) as a function of the applied field.
(c) Pair potential $\Delta$ as a function of normalized field $H/H_\mathrm{c}$. 
(d) Normalized pair-breaking parameter $\Gamma/\Delta_0$ as a function of $(H/H_\mathrm{c})^2$.}
\end{figure}

We now focus on results in the superconducting state. Figure~\ref{fig_local}(a) shows the local differential conductance of one contact as a function of the injection bias voltage $V_\mathrm{inj}$ for different applied magnetic fields $H$ at $T=50~\mathrm{mK}$. The data at zero field show negligible subgap conductance and a gap singularity at $V_\mathrm{inj}\approx220~\mathrm{\mu V}$, consistent with the expected gap $\Delta_0=1.76~k_\mathrm{B}T_\mathrm{c}=230~\mathrm{\mu eV}$. Upon increasing the field, a spin splitting of the density of states quickly develops, which is much larger than the Zeeman splitting due to the applied field. We fit our data with the standard model for high-field tunneling\cite{maki1964b,meservey1975} to extract the normal-state junction conductance $G_\mathrm{inj}$, the spin-polarization $P$ of the tunnel conductance, the pair potential $\Delta$, the orbital pair-breaking parameter $\Gamma$, the spin-orbit scattering strength $b_\mathrm{so}$, and the spin  splitting. The latter appears as an effective magnetic field in the fit, which we denote by $B_\mathrm{fit}$. The spin splitting consists of two parts, the Zeeman splitting due to the applied field $\mu_0 H$, and the exchange splitting induced by the EuS, which we will denote by $B^*=B_\mathrm{fit}-\mu_0 H$, following the literature.\cite{hao1991}

In Fig.~\ref{fig_local}(b), we plot $B^*$ obtained from fitting the spectra of all six junction as a function of the applied field. $B^*$ increases almost linearly at small field, and then saturates for $\mu_0H>0.5~\mathrm{T}$ at  $B_\mathrm{sat}^*=1.75~\mathrm{T}$. Consequently, most of the spin splitting of the density of states results from the exchange field. A similar behavior was found for all samples, with some variation of both the magnitude of $B_\mathrm{sat}^*$ and the applied field where saturation sets in. For comparison, we show the magnetization $M_\mathrm{EuS}$ of the EuS film obtained by SQUID magnetometry in the same plot. As can be seen, the magnetization is fully saturated at a field of a few 10 mT, much below the saturation of the exchange splitting. The reason for the discrepancy of the field dependence of $M_\mathrm{EuS}$ and $B^*$ is not known. In the past, averaging over multi-domain magnetization states of EuS has been assumed as a possible cause of the slow saturation of $B^*$.\cite{hao1991} However, this explanation is at variance with the fast saturation of $M_\mathrm{EuS}$, as well as the small junction size of the order of the dirty-limit coherence length of the aluminum film ($\xi_\mathrm{S}=76~\mathrm{nm}$). Considering the fact that the induced exchange field is the result of spin-active scattering of quasiparticles at the interface between Al and EuS, we assume that a difference between bulk magnetism (as seen by magnetometry) and interface magnetism (as seen by the aluminum) is the cause of the different field scales. Magnetic moments at the interface may deviate from the magnetization direction due to the broken symmetry and different anisotropy at the interface. A further clue supporting this interpretation is the fact that $B_\mathrm{sat}^*$ varies from sample to sample even for similar aluminum film thickness. In our two-step fabrication process, we expect that the interface properties depend sensitively on the Ar ion etching between fabrication steps.

The pair potential $\Delta$, plotted in Fig.~\ref{fig_local}(c), remains almost constant at $\Delta_0\approx 225~\mathrm{\mu eV}$ (dashed line) as a function of applied field up to the critical field. We find $B_\mathrm{sat}^*+\mu_0 H_\mathrm{c}=2.7~\mathrm{T}$ in good agreement with the estimate of the Pauli limiting field $\mu_0 H_\mathrm{p}=\Delta_0/\sqrt{2}\mu_\mathrm{B}=2.75~\mathrm{T}$. The normalized orbital pair breaking parameter $\Gamma/\Delta_0$ is plotted in Fig.~\ref{fig_local}(d) as a function of $(H/H_\mathrm{c})^2$. For a thin film in parallel magnetic field, the expectation is $\Gamma/\Delta_0=(H/H_\mathrm{c})^2/2$ if orbital pair breaking effects dominate.\cite{maki1964a} This assumption is plotted as a dashed line. Indeed, $\Gamma$ follows an $H^2$ dependence at high fields, but with a smaller slope and an additional offset (solid line). From these observations we conclude that the critical field is determined by spin splitting rather than by orbital pair breaking. Below $0.1~\mathrm{T}$, $\Gamma$ increases faster than expected, and the $H^2$ dependence at high fields extrapolates to an offset $\Gamma(H=0)/\Delta_0\approx0.03$. This may indicate additional pair breaking due to magnetic inhomogeneity of the EuS film, or the fringing fields of the iron wires. For the spin-orbit parameter we obtain $b_\mathrm{so}\approx 0.14$ at high fields, much larger than expected for aluminum.\cite{meservey1975}

\begin{figure}[tb]
\includegraphics[width=\columnwidth]{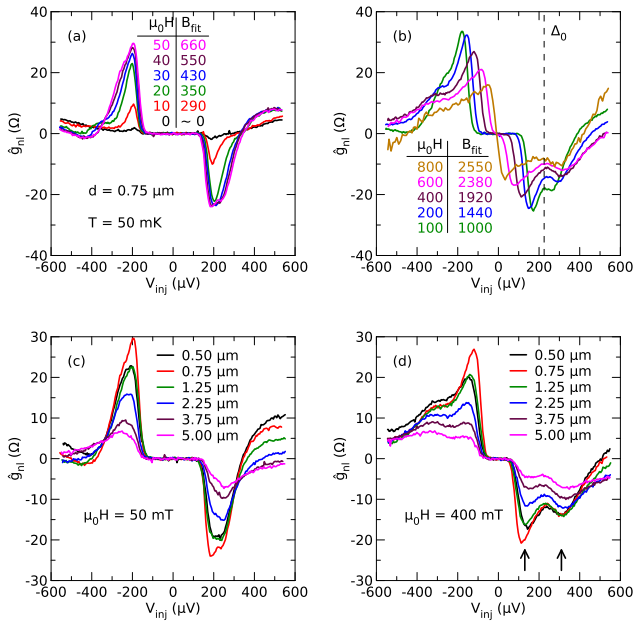}
\caption{\label{fig_nonlocal} 
(color online) Normalized nonlocal differential conductance $\hat{g}_\mathrm{nl}=g_\mathrm{nl}/\left(G_\mathrm{inj}G_\mathrm{det}\right)$ as a function of injector bias $V_\mathrm{inj}$. (a) and (b): Data for different applied magnetic fields $H$ for one pair of contacts. (c) and (d): Data for different contact distances at two different applied fields.}
\end{figure}

In Fig.~\ref{fig_nonlocal}, we focus on the nonlocal differential conductance. Figs.~\ref{fig_nonlocal}(a) and (b) show the normalized nonlocal conductance $\hat{g}_\mathrm{nl}$ as a function of  injector bias $V_\mathrm{inj}$ for different magnetic fields $H$. As in our previous work,\cite{huebler2012b,wolf2013} asymmetric peaks due to spin injection into the spin-split density of states are observed upon increasing the magnetic field, as seen in Fig.~\ref{fig_nonlocal}(a). Due to the increased spin splitting by the exchange field of the EuS, the peaks are clearly visible even at a field as small as $10~\mathrm{mT}$. At larger fields, the peaks broaden due to the increased spin splitting. In contrast to the previous work, however, the peaks actually split into two sub-peaks at small and large bias, as seen in Fig.~\ref{fig_nonlocal}(b). The dashed lines indicate the pair potential. The two sub-peaks essentially follow the spin splitting of the density of states. In  Figs.~\ref{fig_nonlocal}(c) and (d), we show the evolution of the spin signal as a function of contact distance for low and high fields, respectively. In the low-field regime, Fig.~\ref{fig_nonlocal}(c), where a single peak is observed for each bias polarity, the peak uniformly decreases with increasing contact distance. At high fields, Fig.~\ref{fig_nonlocal}(d), the two sub-peaks (indicated by arrows for the positive bias side) decay on different length scales. This can be clearly seen by comparing the data at $d=0.5~\mathrm{\mu m}$ and $5~\mathrm{\mu m}$. At small contact distance, the low-bias peak (at about $V_\mathrm{inj}=140~\mathrm{\mu V}$) is larger, whereas at large distance, the high-bias peak (at about $V_\mathrm{inj}=310~\mathrm{\mu V}$) is larger.

\begin{figure}[bt]
\includegraphics[width=\columnwidth]{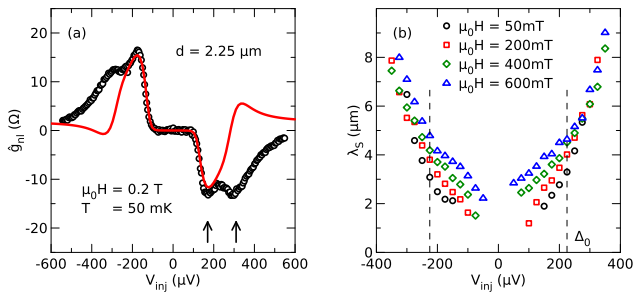}
\caption{\label{fig_relaxation}
(color online) (a) Normalized nonlocal differential conductance $\hat{g}_\mathrm{nl}$ as a function of bias voltage $V_\mathrm{inj}$ for one contact pair and applied field (symbols) and theoretical prediction $\hat{g}_\mathrm{nl}\propto g_\downarrow-g_\uparrow$ (line). (b) Relaxation length $\lambda_\mathrm{S}$ of the spin signal as a function of bias voltage $V_\mathrm{inj}$ for different magnetic fields $H$.}
\end{figure}

Spin injection into the spin-split density of states is proportional to the difference of the conductances for spin up and down, $g_\mathrm{nl}\propto g_\downarrow-g_\uparrow$.\cite{huebler2012b,wolf2013} We can therefore predict the bias dependence of $g_\mathrm{nl}$ from the fits of the local conductance. Fits to this prediction have been successful in describing the data of our previous experiments.\cite{huebler2012b,wolf2013,wolf2014} In Fig.~\ref{fig_relaxation}(a), we compare the bias dependence of the measured spin signal to the predicted signal for one set of parameters. The model predicts a single peak for each bias polarity, restricted to the bias window of the spin splitting. As soon as the bias reaches the upper spin band, the signal should be cancelled by injection of quasiparticles with opposite spin. The observed data are in qualitative contrast to this expectation. Instead of a cancellation, the spin signal actually further increases once the upper spin band is reached and shows a second sub-peak. This observation is systematic for the high-field data.

To analyze the relaxation length of the spin signal, we made exponential fits of $\hat{g}_\mathrm{nl}$ at a given bias as a function of contact distance. In Fig.~\ref{fig_relaxation}(b), the relaxation length  $\lambda_\mathrm{S}$ obtained from these fits is plotted as a function of bias voltage $V_\mathrm{inj}$ for different magnetic fields $H$. $\lambda_\mathrm{S}$ is a few $\mathrm{\mu m}$, similar to samples without EuS film. In general, $\lambda_\mathrm{S}$ increases with bias. At small bias, in the range of the spin splitting, $\lambda_\mathrm{S}$ also increases with magnetic field, again similar to the previous experiments without EuS. At higher bias, where both spin bands contribute to conductance, $\lambda_\mathrm{S}$ becomes nearly independent of the field.

\section{Discussion}

The microscopic explanation of the induced exchange splitting is spin-active scattering at the interface between EuS and Al, which can be expressed in terms of spin-mixing angles.\cite{tokuyasu1988,millis1988} For diffusive systems, a broad distribution of spin-mixing angles is expected. Recently, appropriate boundary conditions for the Usadel equation have been derived.\cite{cottet2009} For a thin superconducting layer on top of a ferromagnetic insulator, the spin-active scattering can be expressed in this model by dimensionless parameters $\gamma_{\phi,i}$. The $\gamma_{\phi,i}$ depend on moments of increasing order of the distribution of spin-mixing angles. The first-order parameter, $\gamma_{\phi,1}$, acts like an effective Zeeman field, whereas the second-order parameter, $\gamma_{\phi,2}$, acts in the same way as pair breaking. The higher-order terms have no apparent analogy. With this analogy, we can express $B_\mathrm{sat}^*$ as $\gamma_{\phi,1}\approx 0.062~(0.049)$ for sample A (B). If we further tentatively attribute the residual pair-breaking strength $\Gamma(H=0)$ to spin-active scattering, this yields $\gamma_{\phi,2}\approx 0.004~(0.002)$. As mentioned in the previous section, we obtain an anomalously large spin-orbit scattering strength $b_\mathrm{so}\approx 0.14$ from the fits. However, if we interpret the normal-state spin-diffusion time as the spin-orbit scattering time (Elliott-Yafet mechanism\cite{elliott1954,yafet1963}), we can estimate $b_\mathrm{so}=\hbar/3\tau_\mathrm{so}\Delta_0=0.024$. This appears much more realistic for aluminum,\cite{meservey1975} and is also similar to literature data for aluminum on EuS.\cite{hao1991} An interesting question to theory is whether fits including higher-order $\gamma_{\phi}$ terms might remove the discrepancy, and provide additional insight into the scattering mechanism at the interface.

For the nonlocal signal, we find that applied fields of 10~mT are sufficient to enable spin injection and transport. The relaxation length of a few microns is similar to what has been found in structures without EuS. At high fields, the spin signal is qualitatively different from the expectation, with increased spin injection instead of cancellation as the upper spin band starts to contribute. In our previous experiments, we have found a high-bias tail of the spin signal in some samples.\cite{huebler2012b,wolf2013} A possible explanation for this behavior are spin flips in combination with fast energy relaxation, as explained in Ref.~\onlinecite{wolf2013}. The same mechanism could be at play here, and might be more pronounced because the density-of-states features are sharper due to the relative weakness of orbital pair-breaking effects. On the other hand, spin-active scattering may lead to the generation of triplet Cooper pairs in the superconductor. This might lead to a qualitatively different spin injection and relaxation behavior. Lacking a quantitative model for either effect, we can only refer this question to theory.

\section{Conclusion}

In conclusion, we have shown spin injection and transport in mesoscopic superconducting aluminum wires with an exchange splitting induced by the ferromagnetic insulator europium sulfide. The salient features observed in the experiment are consistent with the previous literature on spectroscopy and spin transport in high-field superconductivity. Our results show that ferromagnetic insulators are promising materials to control spin transport in superconductors at mesoscopic length scales and to implement spintronics functionality.

\begin{acknowledgments}
We thank W. Belzig, M. Eschrig, and A. Cottet for useful discussions. This work was supported by the Research Network ``Functional Nanostructures'' of the Baden-W\"urttemberg-Stiftung. 
\end{acknowledgments}

\bibliography{lit.bib}

\end{document}